\documentclass[a4paper,oneside,pdftex]{amsart}
\usepackage[T1]{fontenc}
\usepackage[german,english]{babel}
\usepackage{graphicx}
\RequirePackage{amsmath}
\RequirePackage{bm}
\RequirePackage{amssymb}
\RequirePackage{upref}
\RequirePackage{amsthm}
\RequirePackage{enumerate}
\RequirePackage{pb-diagram}
\RequirePackage{amsfonts}
\RequirePackage[mathscr]{eucal}
\RequirePackage{verbatim}
\RequirePackage{xr}
\RequirePackage{graphicx}
\RequirePackage{mathrsfs}
\usepackage{calc}
\usepackage{xspace}
\usepackage{dsfont}

\RequirePackage{amsmath}
\RequirePackage{bm}
\RequirePackage{amssymb}
\RequirePackage{upref}
\RequirePackage{amsthm}
\RequirePackage{enumerate}
\RequirePackage{pb-diagram}
\RequirePackage{amsfonts}
\RequirePackage[mathscr]{eucal}
\RequirePackage{verbatim}
\RequirePackage{xr}
\RequirePackage{graphicx}
\usepackage{calc}
\usepackage{xspace}
\RequirePackage{color}
\RequirePackage{ifthen}
\RequirePackage{fancybox}
\RequirePackage{mathrsfs}

\newcommand{\cf}{cf.\@\xspace}
\newcommand{\resp}{resp.\@\xspace}

\newcommand{\al}{\alpha}
\newcommand{\bet}{\beta}
\newcommand{\ga}{\gamma}
\newcommand{\de}{\delta }
\newcommand{\e}{\epsilon}

\newcommand{\f}{\varphi}
\newcommand{\h}{\eta}

\newcommand{\lam}{\lambda}

\newcommand{\n}{\nu}
\newcommand{\om}{\omega}

\newcommand{\s}{\sigma}
\newcommand{\x}{\xi}

\newcommand{\D}{\varDelta}

\newcommand{\Lam}{\varLambda}
\newcommand{\Om}{\varOmega}

\newcommand{\Si}{\varSigma}

\newcommand{\di}[1]{#1\nobreakdash-\hspace{0pt}dimensional}%\di n

\newcommand{\fv}[2]{#1\hspace{0pt}_{|_{#2}}}

\newcommand{\so}{{\mc S_0}}

\newcommand{\const}{\tup{const}}

\newcommand{\msp[1]}[1]{\mspace{#1mu}}

\newcommand{\R}[1][n+1]{{\protect\mathbb R}^{#1}}

\newcommand{\Cc}{{\protect\mathbb C}}

\newcommand{\N}{{\protect\mathbb N}}

\newcommand{\eR}{\stackrel{\lower1ex \hbox{\rule{6.5pt}{0.5pt}}}{\msp[3]\R[]}}
\newcommand{\eN}{\stackrel{\lower1ex \hbox{\rule{6.5pt}{0.5pt}}}{\msp[1]\N}}
\newcommand{\eO}{\stackrel{\lower1ex \hbox{\rule{6pt}{0.5pt}}}{\msc O}}
\newcommand{\mf}[1]{\mathfrak {#1}}

\DeclareMathOperator{\graph}{graph}

\DeclareMathOperator{\id}{id}

\DeclareMathOperator{\tr}{tr}

\DeclareMathOperator{\diag}{diag}

\DeclareMathOperator{\card}{card}

\newcommand\ra{\rightarrow}

\newcommand\pa{\partial}
\newcommand\pde[2]{\frac {\partial#1}{\partial#2}}
\newcommand\dde[2]{\frac {\de#1}{\de#2}}

\newcommand{\un}{\infty}
\newcommand{\A}{\forall}

\newcommand{\set}[2]{\{\,#1\colon #2\,\}}
\newcommand{\uu}{\cup}

\newcommand{\uuu}{\bigcup}

\newcommand{\uud}{ \stackrel{\lower 1ex \hbox {.}}{\uu}}
\newcommand{\uuud}[1]{ \stackrel{\lower 1ex \hbox {.}}{\uuu_{#1}}}
\newcommand\su{\subset}

\newcommand{\sminus}[1][28]{\raise 0.#1ex\hbox{$\scriptstyle\setminus$}}

\newcommand{\ol}{\overline}

\newcommand{\wed}{\wedge}

\newcommand{\abs}[1]{\lvert#1\rvert}

\newcommand{\norm}[1]{\lVert#1\rVert}

\newcommand{\spd}[2]{\protect\langle #1,#2\protect\rangle}

\newcommand\ch[3]{\varGamma_{#1#2}^#3}
\newcommand\cha[3]{{\bar\varGamma}_{#1#2}^#3}

\newcommand{\riem}[4]{R_{#1#2#3#4}}
\newcommand{\riema}[4]{{\bar R}_{#1#2#3#4}}

\newcommand{\tit}{\textit}

\newcommand{\tup}{\textup}% text upright

\newcommand{\mc}{\protect\mathcal}
\newcommand{\msc}{\protect\mathscr}

\providecommand{\bysame}{\makebox[3em]{\hrulefill}\thinspace}

\newcommand{\cfd}{\hat{\msc S'}(\mf a^*\times B)}
\newcommand{\cfs}{\hat{\msc S}(\mf a^*\times B)}

\newcommand{\cq}[1]{\glqq{#1}\grqq\,}

\newcommand{\bt}{\begin{thm}}
\newcommand{\bl}{\begin{lem}}
\newcommand{\bc}{\begin{cor}}
\newcommand{\bd}{\begin{definition}}
\newcommand{\bpp}{\begin{prop}}
\newcommand{\br}{\begin{rem}}
\newcommand{\bn}{\begin{note}}
\newcommand{\be}{\begin{ex}}
\newcommand{\bes}{\begin{exs}}
\newcommand{\bb}{\begin{example}}
\newcommand{\bbs}{\begin{examples}}
\newcommand{\ba}{\begin{axiom}}
\newcommand{\bas}{\begin{assumption}}

\newcommand{\et}{\end{thm}}
\newcommand{\el}{\end{lem}}
\newcommand{\ec}{\end{cor}}
\newcommand{\ed}{\end{definition}}
\newcommand{\epp}{\end{prop}}
\newcommand{\er}{\end{rem}}
\newcommand{\en}{\end{note}}
\newcommand{\ee}{\end{ex}}
\newcommand{\ees}{\end{exs}}
\newcommand{\eb}{\end{example}}
\newcommand{\ebs}{\end{examples}}
\newcommand{\ea}{\end{axiom}}
\newcommand{\eas}{\end{assumption}}

\newcommand{\bp}{\begin{proof}}
\newcommand{\ep}{\end{proof}}
\newcommand{\eps}{\renewcommand{\qed}{}\end{proof}}

\newcommand{\bal}{\begin{align}}

\newcommand{\bi}[1][1.]{\begin{enumerate}[\upshape #1]}
\newcommand{\bia}[1][(1)]{\begin{enumerate}[\upshape #1]}
\newcommand{\bin}[1][1]{\begin{enumerate}[\upshape\bfseries #1]}
\newcommand{\bir}[1][(i)]{\begin{enumerate}[\upshape #1]}
\newcommand{\bic}[1][(i)]{\begin{enumerate}[\upshape\hspace{2\cma}#1]}
\newcommand{\bis}[2][1.]{\begin{enumerate}[\upshape\hspace{#2\parindent}#1]}
\newcommand{\ei}{\end{enumerate}}

\newcommand\ndots{\raise 0.47ex \hbox {,}\hskip0.06em\cdots %
     \raise 0.47ex \hbox {,}\hskip0.06em} 

\newcommand{\q}{\quad}
\newcommand{\qq}{\qquad}

\newcommand{\hp}{\hphantom}

\newcommand\nd{\noindent}

\newskip\Csmallskipamount                                                
\Csmallskipamount=\smallskipamount
\newskip\Cmedskipamount
\Cmedskipamount=\medskipamount
\newskip\Cbigskipamount
\Cbigskipamount=\bigskipamount

\newcommand\cvs{\vspace\Csmallskipamount}   
\newcommand\cvm{\vspace\Cmedskipamount}

\newskip\csa
\csa=\smallskipamount

\newskip\cma
\cma=\medskipamount

\newskip\cba
\cba=\bigskipamount

\newdimen\spt
\newcommand\citem{\cvs\advance\itemno by
1{(\romannumeral\the\itemno})\hskip3pt}
\newcommand{\bitem}{\cvm\nd\advance\itemno by
1{\bf\the\itemno}\hspace{\cma}}

\newcount\itemno
\newcommand{\las}[1]{\label{S:#1}}

\newcommand{\lae}[1]{\label{E:#1}}

\newcommand{\lad}[1]{\label{D:#1}}

\newcommand{\lar}[1]{\label{R:#1}}

\newcommand{\rs}[1]{Section~\ref{S:#1}}

\newcommand{\re}[1]{\eqref{E:#1}}

\newcommand{\frr}[1]{Remark~\ref{R:#1} on page~\tup{\pageref{R:#1}}}
\newcommand{\frd}[1]{Definition~\ref{D:#1} on page~\tup{\pageref{D:#1}}}
\newcommand{\fre}[1]{\eqref{E:#1} on page~\tup{\pageref{E:#1}}}

\newskip\thmskip
\thmskip=\parindent

\newskip\hsk
\newenvironment{hinw}{\labelsep=0pt\begin{list}{}{\labelsep=0pt\itemindent=0pt\labelwidth=0pt\leftmargin=\parindent\rightmargin=0pt\partopsep=\cba}%
\item\it\nopagebreak\nopagebreak}%
{\end{list}}

\newcommand\bh{\begin{hinw}}
\newcommand{\eh}{\end{hinw}}

\newtheoremstyle{normal}% name
  {\cba}%      Space above, empty = `usual value'
  {\cba}%      Space below
  {}% Body font
  {\thmskip}%Indent amount (empty = no indent, \parindent = para indent)
  {\bfseries}% Thm head font
  {.}%        Punctuation after thm head
  {\hsk}%     Space after thm head: " " = normal interword space;
  {}% Thm head spec

\newtheoremstyle{abschnitt}% name
  {\cba}%      Space above, empty = `usual value'
  {\cba}%      Space below
  {}% Body font
  {\thmskip}% Indent amount (empty = no indent, \parindent = para indent)
  {\bfseries}% Thm head font
  {.}%        Punctuation after thm head
  {\hsk}%     Space after thm head: " " = normal interword space;
  {}% Thm head spec

\newtheoremstyle{italic}% name
  {\cba}%      Space above, empty = `usual value'
  {\cba}%      Space below
  {\itshape}% Body font
  {\thmskip}%  Indent amount (empty = no indent, \parindent = para indent)
  {\bfseries}% Thm head font
  {.}%        Punctuation after thm head
  {\hsk}%     Space after thm head: " " = normal interword space;
  {}% Thm head spec

\newtheoremstyle{aufgaben}% name
  {\cba}%      Space above, empty = `usual value'
  {\cba}%      Space below
  {}% Body font
  {}%         Indent amount (empty = no indent, \parindent = para indent)
  {\normalsize\bfseries}% Thm head font
  {.}%        Punctuation after thm head
  {\hsk}%     Space after thm head: " " = normal interword space;
  {}% Thm head spec

\newtheoremstyle{break}% name
  {\cba}%      Space above, empty = `usual value'
  {\cba}%      Space below
  {\itshape}% Body font
  {}%         Indent amount (empty = no indent, \parindent = para indent)
  {\bfseries}% Thm head font
  {.}%        Punctuation after thm head
  {\newline}% Space after thm head: \newline = linebreak
  {}%         Thm head spec

\theoremstyle{italic}
\newtheorem{thm}[subsection]{Theorem}
\newtheorem{lem}[subsection]{Lemma}
\newtheorem{prop}[subsection]{Proposition}
\newtheorem{cor}[subsection]{Corollary}

\theoremstyle{normal}
\newtheorem{rem}[subsection]{Remark}
\newtheorem{definition}[subsection]{Definition}
\newtheorem{example}[subsection]{Example}
\newtheorem{examples}[subsection]{Examples}
\newtheorem{ex}[subsection]{Exercise}
\newtheorem{note}[subsection]{}
\newtheorem{axiom}[subsection]{Axiom}
\newtheorem{assumption}[subsection]{Assumption}

\theoremstyle{aufgaben}
\newtheorem{exs}[subsection]{Exercises}

\numberwithin{equation}{section}
\numberwithin{figure}{section}

\newenvironment{textequation}[1][0.8]
{\begin{equation}
\begin{aligned}
\begin{minipage}{#1\linewidth}}
{\end{minipage}
\end{aligned}
\end{equation}
\ignorespacesafterend}

\newcommand{\btext}{\begin{textequation}}
\newcommand{\etext}{\end{textequation}}

\def\hinweis{\@startsection{subsection}{2}%
 \z@{0.7\linespacing\@plus 0.5\linespacing}{0.7\linespacing}%
{\normalfont\itshape\indent}}

\newcounter{hours}\newcounter{minutes}
\newcommand{\printtime}{%
\setcounter{hours}{\time/60}%
\setcounter{minutes}{\time-\value{hours}*60}%
\ifthenelse{\value{minutes}<10}{\thehours :0\theminutes}{\thehours:\theminutes}}

\makeatletter
\RequirePackage{color}
\newcommand{\ann}[1]{\renewcommand{\@makefnmark}{\mbox{$^{\color{red}{\@thefnmark}}$}}%
\footnote {#1}}
\makeatother

\newcommand{\ind}[1]{#1 \relax}

\newcommand{\indexs}[1]{\relax}
\newcommand{\inds}[1]{#1\relax}
\newcommand{\indss}[1]{\relax}

\RequirePackage{upref}
\RequirePackage{amsthm}
\RequirePackage{enumerate}%\begin{enumerate}[(i)]
\usepackage[mathscr]{eucal}
\usepackage{xr-hyper}

\usepackage{calc}

\newlength{\oddsidemarginlength}
\newlength{\topmarginlength}

\newcounter{numberoflines}
\newcounter{tempcc}
\oddsidemargin=\oddsidemarginlength
\evensidemargin=\oddsidemargin
\topmargin=\topmarginlength
\usepackage[colorlinks=true,linkcolor=blue,citecolor=blue,urlcolor=blue]{hyperref}  

\begin{document}

\flushbottom

%\larger[1]
%\frontmatter

\title[Quantization of the Hamilton equations]{The quantization of gravity: Quantization of the Hamilton equations}

% author one information
\author{Claus Gerhardt}
\address{Ruprecht-Karls-Universit\"at, Institut f\"ur Angewandte Mathematik,
Im Neuenheimer Feld 205, 69120 Heidelberg, Germany}
%\curraddr{}
\email{\href{mailto:gerhardt@math.uni-heidelberg.de}{gerhardt@math.uni-heidelberg.de}}
\urladdr{\href{http://www.math.uni-heidelberg.de/studinfo/gerhardt/}{http://www.math.uni-heidelberg.de/studinfo/gerhardt/}}
%\thanks{This work was supported by the DFG}

% author two information
%\author{}
%\address{}
%\curraddr{}
%\email{}
%\thanks{}
%
%\subjclass[2000]{35J60, 53C21, 53C44, 53C50, 58J05}
%\keywords{Lorentzian manifold, mass, cosmological spacetime, general relativity, inverse mean curvature flow, ARW spacetimes}


\date{\today}
%
% at present the "communicated by" line appears only in ERA and PROC
%\commby{}

%\dedicatory{}

\begin{abstract} 
We quantize the Hamilton equations instead of the Hamilton condition. The resulting equation has the simple form $-\D u=0$ in a fiber bundle, where the Laplacian is the Laplacian of the Wheeler-DeWitt metric provided $n\not=4$. Using then separation of variables  the solutions $u$ can be expressed as products of temporal and spatial eigenfunctions, where the spatial eigenfunctions are eigenfunctions of the Laplacian in the symmetric space $SL(n,\R[])/SO(n)$. Since one can define a Schwartz space and tempered distributions in $SL(n,\R[])/SO(n)$  as well as  a Fourier transform,  Fourier quantization can be applied such that the spatial eigenfunctions are transformed to Dirac measures and the spatial Laplacian to a multiplication operator.
\end{abstract}

\maketitle

\tableofcontents

\setcounter{section}{0}
\section{Introduction}
General relativity is a Lagrangian theory, i.e., the Einstein equations are derived as the Euler-Lagrange equation of the Einstein-Hilbert functional
\begin{equation}
\int_N(\bar R-2\Lam),
\end{equation}
where $N=N^{n+1}$, $n\ge 3$, is a globally hyperbolic Lorentzian manifold, $\bar R$ the scalar curvature and $\Lam$ a cosmological constant. We also omitted the integration density in the integral. In order to apply a Hamiltonian description of general relativity, one usually defines a time function $x^0$ and considers the foliation of $N$ given by the slices
\begin{equation}
M(t)=\{x^0=t\}.
\end{equation}
We may, without loss of generality, assume that the spacetime metric splits
\begin{equation}
d\bar s^2=-w^2(dx^0)^2+g_{ij}(x^0,x)dx^idx^j,
\end{equation}
\cf \cite[Theorem 3.2]{cg:qgravity}. Then, the Einstein equations also split into a tangential part
\begin{equation}
G_{ij}+\Lam g_{ij}=0
\end{equation}
and a normal part
\begin{equation}
G_{\al\bet}\nu^\al\nu^\bet-\Lam=0,
\end{equation}
where the naming refers to the given foliation. For the tangential Einstein equations one can define equivalent Hamilton equations due to the groundbreaking paper by Arnowitt, Deser and Misner \cite{adm:old}. The normal Einstein equations can be expressed by the so-called Hamilton condition
\begin{equation}\lae{1.6}
\mc H=0,
\end{equation}
where $\mc H$ is the Hamiltonian used in defining the Hamilton equations. In the canonical quantization of gravity the Hamiltonian is transformed  to a partial differential operator of hyperbolic type $\hat{\mc H}$ and the possible quantum solutions of gravity are supposed to satisfy the so-called Wheeler-DeWitt equation
\begin{equation}\lae{1.7}
\hat{\mc H}u=0
\end{equation}
in an appropriate setting, i.e., only the Hamilton condition \re{1.6} has been quantized, or equivalently, the normal Einstein equation, while the tangential Einstein equations have been ignored.

In \cite{cg:qgravity} we solved the equation \re{1.7} in a fiber bundle $E$ with base space $\so$,
\begin{equation}
\so=\{x^0=0\}\equiv M(0),
\end{equation}
and fibers $F(x)$, $x\in\so$,
\begin{equation}
F(x)\su T^{0,2}_x(\so),
\end{equation}
the elements of which are the positive definite symmetric tensors of order two, the Riemannian metrics in $\so$. The hyperbolic operator $\hat{\mc H}$ is then expressed in the form
\begin{equation}\lae{1.10}
\hat{\mc H}=-\D-(R-2\Lam)\f,
\end{equation}
where $\D$ is the Laplacian of the Wheeler-DeWitt metric given in the fibers, $R$ the scalar curvature of the metrics $g_{ij}(x)\in F(x)$, and $\f$ is defined by
\begin{equation}\lae{1.11}
\f^2=\frac{\det g_{ij}}{\det\chi_{ij}},
\end{equation}
where $\chi_{ij}$ is a fixed metric in $\so$ such that instead of densities we are considering functions. The Wheeler-DeWitt equation could be solved in $E$ but only as an abstract hyperbolic equation. The solutions could not be split in corresponding spatial and temporal eigenfunctions.

Therefore, we  discarded the Wheeler-DeWitt equation in \cite{cg:qgravity2b}, see also \cite[Chapter 1]{cg:qgravity-book},  and looked at the evolution equations given by the second Hamilton equation. The left-hand side, a time derivative, we  replaced with the help of the  Poisson brackets. On the right-hand side we implemented the Hamilton condition, equation \re{1.6}. After canonical quantization the Poisson brackets  became a commutator and we applied both sides of the equation to smooth functions with compact support defined in the fiber bundle. The resulting equation we evaluated for a particular metric which we considered important to the problem and then obtained a hyperbolic equation in the base space, which happened to be identical to the Wheeler-DeWitt equation obtained as a result of a canonical quantization of a Friedman universe, if we only looked at functions that did not depend on $x$ but only on the scale factor, which now acted as a time variable. Evidently, this result can not be regarded as the solution to the problem of quantizing gravity in a general setting.

 The underlying mathematical reason for the difficulty was the presence of the term $R$ in the quantized equation, which prevents the application of separation of variables, since the metrics $g_{ij}$ are the spatial variables. In this paper we overcome this difficulty by quantizing the Hamilton equations without alterations, i.e., we completely discard the Hamilton condition. From a logical point of view this approach  is as justified as the prior procedure by quantizing only the normal Einstein equation and discarding the tangential Einstein equations---despite the fact that the tangential Einstein equations are equivalent to the Hamilton equations. This equivalence is considered to be an essential prerequisite for canonical quantization, which is the quantization of the Hamilton equations.
 
 During quantization the transformed Hamiltonian is acting on smooth functions $u$ which are only defined in the fibers, i.e., they only depend on the metrics $g_{ij}$ and not explicitly on $x\in\so$. As result we obtain the equation
 \begin{equation}\lae{1.12}
-\D u=0
\end{equation}
in $E$, where the Laplacian is the Laplacian in \re{1.10}. The lower order terms of $\hat{\mc H}$ 
\begin{equation}
(R-2\Lam)\f
\end{equation}
present on both sides of the equation cancel each other. However, the equation \re{1.12} is only valid provided $n\not=4$, since the resulting equation actually looks like
\begin{equation}
-(\frac n2-2)\D u=0.
\end{equation}
This restriction seems to be acceptable, since $n$ is the dimension of the base space $\so$ which, by general consent, is assumed to be $n=3$. The fibers add additional dimensions to the quantized problem, namely,
\begin{equation}
\dim F=\frac {n(n+1)}2\equiv m+1.
\end{equation}
The fiber metric, the Wheeler-DeWitt metric, which is responsible for the Laplacian in \re{1.12} can be expressed in the form
\begin{equation}
ds^2=-\frac{16(n-1)}n dt^2+\f G_{AB}d\xi^Ad\xi^B,
\end{equation}
where the coordinate system is
\begin{equation}
(\xi^a)= (\xi^0,\xi^A)\equiv (t,\xi^A).
\end{equation}
The $(\xi^A)$, $1\le A\le m$, are coordinates for the hypersurface
\begin{equation}\lae{1.18}
M\equiv M(x)=\{(g_{ij}):t^4=\det g_{ij}(x)=1,\A\, x\in\so\}.
\end{equation}
We also assume that $\so=\R[n]$ and that  the metric $\chi_{ij}$ in \re{1.11} is the Euclidean metric $\de_{ij}$. It is well-known that $M$ is a symmetric space
\begin{equation}
M=SL(n,\R[])/SO(n)\equiv G/K.
\end{equation}
It is also easily verified that the induced metric of $M$ in $E$ is identical to the Riemannian metric of the coset space $G/K$.

Now, we are in a position to use separation of variables, namely, we write a solution of \re{1.12} in the form
\begin{equation}
u=w(t)v(\xi^A),
\end{equation}
where $v$ is a spatial eigenfunction of the induced Laplacian of $M$
\begin{equation}
-\D_Mv\equiv -\D v=(\abs\lam^2+\abs\rho^2)v
\end{equation}
and $w$ is a temporal eigenfunction satisfying the ODE
\begin{equation}\lae{1.22}
\Ddot w+m t^{-1}\dot w+\mu_0 t^{-2}w=0
\end{equation}
with
\begin{equation}
\mu_0=\frac{16(n-1)}n(\abs \lam^2+\abs\rho^2).
\end{equation}

The eigenfunctions of the Laplacian in $G/K$ are well-known and we choose the kernel of the Fourier transform in $G/K$ in order to define the eigenfunctions. This choice also allows us to use Fourier quantization similar to the Euclidean case such that the eigenfunctions are transformed to Dirac measures and the Laplacian to a multiplication operator in Fourier space.

Here is a more detailed overview of the main results. Let $NAK$ be an Iwasawa decomposition of $G$ and 
\begin{equation}
\mf g=\mf n+\mf a+\mf k
\end{equation}
be the corresponding direct sum of the Lie algebras. Let $\mf a^*$ be the dual space of $\mf a$, then the Fourier kernel is defined by the eigenfunctions 
\begin{equation}
e_{\lam,b}(x)=e^{(i\lam+\rho)\log A(x,b)}
\end{equation}
with $\lam\in \mf a^*$, $x=gK\in G/K$, and $b\in B$, where $B$ is the Furstenberg bound\-ary, see \rs{4} and \rs{5} for detailed definitions and references. We then pick a particular $b_0\in B$ and use $e_{\lam,b_0}$ as eigenfunctions of $-\D$
\begin{equation}
-\D e_{\lam,b_0}=(\abs\lam ^2+\abs\rho^2) e_{\lam,b_0}.
\end{equation}
The Fourier transform of $e_{\lam,b_0}$ is
\begin{equation}
\hat e_{\lam,b_0}=\de_\lam\otimes \de_{b_0}
\end{equation}
and of $-\D f$
\begin{equation}
\mc F(-\D f)=(\abs\lam ^2+\abs\rho^2)\hat f,\qq \lam\in \mf a^*, f\in \msc S(G/K).
\end{equation}
The elementary gravitons correspond to special characters in $\mf a^*$, namely,
\begin{equation}
\al_{ij},\q 1\le i<j\le n,
\end{equation}
for the off-diagonal gravitons and
\begin{equation}
\al_i,\q 1\le i\le n-1
\end{equation}
for the diagonal gravitons. Note, that only $(n-1)$ diagonal elements $g_{ii}$ can be freely chosen because of the condition \re{1.18}.

To define the temporal eigenfunctions, we shall here only consider the case $3\le n\le 16$, then all temporal eigenfunctions are generated by the two real eigenfunctions contained in
\begin{equation}
w(t)=t^{-\frac{m-1}2}e^{i\mu\log t},
\end{equation}
where $\mu>0$ is chosen appropriately. These eigenfunctions become unbounded if the big bang (t=0) is approached and they vanish if $t$ goes to infinity.

\section{Definitions and notations}
The main objective of this section is to state the equations of Gau{\ss}, Codazzi,
and Weingarten for spacelike hypersurfaces $M$ in a \di {(n+1)} Lorentzian
manifold
$N$.  Geometric quantities in $N$ will be denoted by
$(\bar g_{ \al \bet}),(\riema  \al \bet \ga \de)$, etc., and those in $M$ by $(g_{ij}), 
(\riem ijkl)$, etc.. Greek indices range from $0$ to $n$ and Latin from $1$ to $n$;
the summation convention is always used. Generic coordinate systems in $N$ resp.
$M$ will be denoted by $(x^ \al)$ \resp $(\x^i)$. Covariant differentiation will
simply be indicated by indices, only in case of possible ambiguity they will be
preceded by a semicolon, i.e., for a function $u$ in $N$, $(u_ \al)$ will be the
gradient and
$(u_{ \al \bet})$ the Hessian, but e.g., the covariant derivative of the curvature
tensor will be abbreviated by $\riema  \al \bet \ga{ \de;\e}$. We also point out that
\begin{equation}
\riema  \al \bet \ga{ \de;i}=\riema  \al \bet \ga{ \de;\e}x_i^\e
\end{equation}
with obvious generalizations to other quantities.

Let $M$ be a \tit{spacelike} hypersurface, i.e., the induced metric is Riemannian,
with a differentiable normal $\n$ which is timelike.

In local coordinates, $(x^ \al)$ and $(\x^i)$, the geometric quantities of the
spacelike hypersurface $M$ are connected through the following equations
\begin{equation}\lae{01.2}
x_{ij}^ \al= h_{ij}\n^ \al
\end{equation}
the so-called \tit{Gau{\ss} formula}. Here, and also in the sequel, a covariant
derivative is always a \tit{full} tensor, i.e.

\begin{equation}
x_{ij}^ \al=x_{,ij}^ \al-\ch ijk x_k^ \al+ \cha  \bet \ga \al x_i^ \bet x_j^ \ga.
\end{equation}
The comma indicates ordinary partial derivatives.

In this implicit definition the \tit{second fundamental form} $(h_{ij})$ is taken
with respect to $\n$.

The second equation is the \tit{Weingarten equation}
\begin{equation}
\n_i^ \al=h_i^k x_k^ \al,
\end{equation}
where we remember that $\n_i^ \al$ is a full tensor.

Finally, we have the \tit{Codazzi equation}
\begin{equation}
h_{ij;k}-h_{ik;j}=\riema \al \bet \ga \de\n^ \al x_i^ \bet x_j^ \ga x_k^ \de
\end{equation}
and the \tit{Gau{\ss} equation}
\begin{equation}
\riem ijkl=- \{h_{ik}h_{jl}-h_{il}h_{jk}\} + \riema  \al \bet\ga \de x_i^ \al x_j^ \bet
x_k^ \ga x_l^ \de.
\end{equation}

Now, let us assume that $N$ is a globally hyperbolic Lorentzian manifold with a
 Cauchy surface. 
$N$ is then a topological product $I\times \mc S_0$, where $I$ is an open interval,
$\mc S_0$ is a  Riemannian manifold, and there exists a Gaussian coordinate
system
$(x^ \al)$, such that the metric in $N$ has the form 
\begin{equation}\lae{01.7}
d\bar s_N^2=e^{2\psi}\{-{dx^0}^2+\s_{ij}(x^0,x)dx^idx^j\},
\end{equation}
where $\s_{ij}$ is a Riemannian metric, $\psi$ a function on $N$, and $x$ an
abbreviation for the spacelike components $(x^i)$. 
We also assume that
the coordinate system is \tit{future oriented}, i.e., the time coordinate $x^0$
increases on future directed curves. Hence, the \tit{contravariant} timelike
vector $(\x^ \al)=(1,0,\dotsc,0)$ is future directed as is its \tit{covariant} version
$(\x_ \al)=e^{2\psi}(-1,0,\dotsc,0)$.

Let $M=\graph \fv u\so$ be a spacelike hypersurface
\begin{equation}
M=\set{(x^0,x)}{x^0=u(x),\,x\in\mc S_0},
\end{equation}
then the induced metric has the form
\begin{equation}
g_{ij}=e^{2\psi}\{-u_iu_j+\s_{ij}\}
\end{equation}
where $\s_{ij}$ is evaluated at $(u,x)$, and its inverse $(g^{ij})=(g_{ij})^{-1}$ can
be expressed as
\begin{equation}\lae{01.10}
g^{ij}=e^{-2\psi}\{\s^{ij}+\frac{u^i}{v}\frac{u^j}{v}\},
\end{equation}
where $(\s^{ij})=(\s_{ij})^{-1}$ and
\begin{equation}\lae{01.11}
\begin{aligned}
u^i&=\s^{ij}u_j\\
v^2&=1-\s^{ij}u_iu_j\equiv 1-\abs{Du}^2.
\end{aligned}
\end{equation}
Hence, $\graph u$ is spacelike if and only if $\abs{Du}<1$.

The covariant form of a normal vector of a graph looks like
\begin{equation}
(\n_ \al)=\pm v^{-1}e^{\psi}(1, -u_i).
\end{equation}
and the contravariant version is
\begin{equation}
(\n^ \al)=\mp v^{-1}e^{-\psi}(1, u^i).
\end{equation}
Thus, we have
\br Let $M$ be spacelike graph in a future oriented coordinate system. Then the
contravariant future directed normal vector has the form
\begin{equation}
(\n^ \al)=v^{-1}e^{-\psi}(1, u^i)
\end{equation}
and the past directed
\begin{equation}\lae{01.15}
(\n^ \al)=-v^{-1}e^{-\psi}(1, u^i).
\end{equation}
\er

In the Gau{\ss} formula \re{01.2} we are free to choose the future or past directed
normal, but we stipulate that we always use the past directed normal.
Look at the component $ \al=0$ in \re{01.2} and obtain in view of \re{01.15}

\begin{equation}\lae{01.16}
e^{-\psi}v^{-1}h_{ij}=-u_{ij}- \cha 000\mspace{1mu}u_iu_j- \cha 0j0
\mspace{1mu}u_i- \cha 0i0\mspace{1mu}u_j- \cha ij0.
\end{equation}
Here, the covariant derivatives are taken with respect to the induced metric of
$M$, and
\begin{equation}
-\cha ij0=e^{-\psi}\bar h_{ij},
\end{equation}
where $(\bar h_{ij})$ is the second fundamental form of the hypersurfaces
$\{x^0=\const\}$.

An easy calculation shows
\begin{equation}
\bar h_{ij}e^{-\psi}=-\tfrac{1}{2}\dot\s_{ij} -\dot\psi\s_{ij},
\end{equation}
where the dot indicates differentiation with respect to $x^0$.

\section{The Hamiltonian approach to general relativity}\las{2}

The Einstein equations with a cosmological constant $\Lam$ in a Lorentzian manifold $N=N^{n=1}$, $n\ge 3$, with metric $\bar g_{\al\bet}$, $ 0\le \al,\bet\le n$, are the Euler-Lagrange equations of the functional
\begin{equation}\lae{2.1}
J=\int_N(\bar R-2\Lam),
\end{equation}
where $\bar R$ is the scalar curvature of the metric and where we omitted the density $\sqrt{\abs{\bar g}}$. The Euler-Lagrange equations are
\begin{equation}
G_{\al\bet}+\Lam \bar g_{\al\bet}=0,
\end{equation}
where $G_{\al\bet}$ is the Einstein tensor. We proved in \cite[Theorem 3.2]{cg:qgravity}, see also \cite[Theorem 1.3.2]{cg:qgravity-book}, that it suffices to consider only metrics that split, i.e., metrics that are of the form
\begin{equation}
d\bar s^2=-w^2 (dx^0)^2+g_{ij}(x^0,x)dx^idx^j,
\end{equation}
where $(x^i)$ are spatial coordinates, $x^0$ is a time coordinate, $g_{ij}$ are Riemannian metrics defined on the slices
\begin{equation}
M(t)=\{x^0=t\},\qq t\in (a,b)
\end{equation}
and 
\begin{equation}
0<w=w(x^0,x)
\end{equation}
is an arbitrary smooth function in $N$.

A stationary metric in that restricted class is also stationary with respect to arbitrary compact variations and, hence, satisfies the full Einstein equations.

Following Arnowitt, Deser and Misner \cite{adm:old} the functional in \re{2.1} can be expressed in the form
\begin{equation}\lae{1.3.37}
J=\int_a^b\int_\Om\{\abs A^2-H^2+R-2\Lam\}w \sqrt g,
\end{equation}
\cf \cite[equ. (1.3.37)]{cg:qgravity-book}, where
\begin{equation}
\abs A^2=h^{ij}h_{ij}
\end{equation}
is the square of the second fundamental form of the slices $M(t)$
\begin{equation}\lae{2.8}
h_{ij}=-\tfrac12w^{-1}\dot g_{ij},
\end{equation}
$H^2$ is the square of the mean curvature
\begin{equation}
H=g^{ij}h_{ij},
\end{equation}
$R$ the scalar curvature of the slices $M(t)$, the interval $(a,b)$ is compactly contained in
\begin{equation}
I=x^0(N)
\end{equation}
and $\Om$ is a bounded open subset  of the fixed slice
\begin{equation}
\so\equiv M(0),
\end{equation}
where we assume
\begin{equation}
0\in I.
\end{equation}
Here, we also assume $N$ to be  globally hyperbolic such that there exists a global time function and $N$ can be written as a topological product
\begin{equation}
N=I\times \so.
\end{equation}

Let $F=F(h_{ij})$ be the \ind{scalar curvature operator} 
\begin{equation}
F=\tfrac12(H^2-\abs A^2)
\end{equation}
and let 
\begin{equation}
F^{ij,kl}=g^{ij}g^{kl}-\tfrac12\{g^{ik}g^{jl}+g^{il}g^{jk}\}
\end{equation}
be its Hessian, then\indss{$F^{ij,kl}$}
\begin{equation}\lae{3.3.14}
F^{ij,kl}h_{ij}h_{kl}=2F=H^2-\abs A^2
\end{equation}
and
\begin{equation}\lae{3.3.15}
F^{ij}=F^{ij,kl}h_{kl}=Hg^{ij}-h^{ij}.
\end{equation}
In physics\indss{$G^{ij,kl}$}
\begin{equation}\lae{3.3.16}
G^{ij,kl}=-F^{ij,kl}
\end{equation}
is known as the DeWitt metric, or more precisely, a conformal metric, where the conformal factor is even a density, is known as the DeWitt metric, but we prefer the above definition.

Combining \re{2.8} and \re{3.3.14} $J$ can be expressed in the form 
\begin{equation}\lae{3.3.17}
J=\int_a^b\int_\Om\{\tfrac14 G^{ij,kl}\dot g_{ij}\dot g_{kl}w^{-2}+(R-2\Lam)\}w\sqrt g.
\end{equation}
The Lagrangian density $\mc L$ is a regular Lagrangian with respect to the variables $g_{ij}$. Define the \ind{conjugate momenta}\indexs{$\pi^{ij}$}
\begin{equation}\lae{3.3.18}
\begin{aligned}
\pi^{ij}=\frac{\pa\mc L}{\pa \dot g_{ij}}&=\tfrac12 G^{ij,kl}\dot g_{kl}w^{-1}\sqrt g\\
&= -G^{ij,kl}h_{kl}\sqrt g
\end{aligned}
\end{equation}
and the \ind{Hamiltonian density}  
\begin{equation}
\begin{aligned}
\mc H&=\pi^{ij}\dot g_{ij}-\mc L\\
&=\frac1{\sqrt g}wG_{ij,kl}\pi^{ij}\pi^{kl}-(R-2\Lam)w\sqrt g,
\end{aligned}
\end{equation}
where\indss{$G_{ij,kl}$}
\begin{equation}
G_{ij,kl}=\tfrac12\{g_{ik}g_{jk}+g_{il}g_{jk}\}-\tfrac1{n-1}g_{ij}g_{kl}
\end{equation}
is the inverse of $G^{ij,kl}$.

Since the Lagrangian is regular with respect to the variables $g_{ij}$, the tangential Einstein equations
\begin{equation}
G_{ij}+\Lam g_{ij}=0
\end{equation}
are equivalent to the Hamilton equations
\begin{equation}
\dot g_{ij}=\frac{\de\mc H}{\de \pi^{ij}}
\end{equation}
and
\begin{equation}
\dot\pi^{ij}=-\frac{\de\mc H}{\de g_{ij}},
\end{equation}
where the differentials on the right-hand side of these equations are variational or functional derivatives.

The mixed Einstein equations vanish
\begin{equation}
G_{0j}+\Lam \bar g_{0j}=0,\q 1\le j\le n,
\end{equation}
and the normal equation
\begin{equation}
G_{\al\bet}\nu^\al\nu^\bet-\Lam=0
\end{equation}
is equivalent to
\begin{equation}
\abs A^2-H^2=R-2\Lam,
\end{equation}
\cf \cite[equ. 1.1.43]{cg:cp}, which in turn is equivalent to
\begin{equation}
\mc H=0,
\end{equation}
which is also known as the Hamilton condition.

We define the \ind{Poisson brackets}
\begin{equation}
\{u,v\}=\dde u{g_{kl}}\dde v{\pi^{kl}}-\dde u{\pi^{kl}}\dde v{g_{kl}}
\end{equation}
and obtain
\begin{equation}\lae{3.4.2.1}
\{g_{ij},\pi^{kl}\}=\de^{kl}_{ij},
\end{equation}
where\indss{$\de^{kl}_{ij}$} 
\begin{equation}
\de^{kl}_{ij}=\tfrac12\{\de^k_i\de^l_j+\de^l_i\de^k_j\}.
\end{equation}
Then, the second Hamilton equation can also be expressed as
\begin{equation}\lae{3.3.32}
\dot \pi^{ij}=\{\pi^{ij},\mc H\}.
\end{equation}
In the next section we want to quantize the Hamilton equations or, more precisely,
\begin{equation}\lae{2.34}
\begin{aligned}
g_{ij}\{\pi^{ij},\mc H\}&=-g_{ij}\frac{\de\mc H}{\de g_{ij}}\\
&=(\frac n2-2)(\abs A^2-H^2)w\sqrt g+\frac n2(R-2\Lam)w\sqrt g\\
 &\qq \qq -Rw\sqrt g-(n-1)\tilde\D w\sqrt g,
\end{aligned}
\end{equation}
\cf \cite[equ. (1.3.64), (1.3.65)]{cg:qgravity-book}, where $\tilde\D$ is the Laplacian with respect to the metric $g_{ij}(t,\cdot)$. 

\section{The quantization}\las{3}
For the quantization of the Hamiltonian setting we first replace all densities by tensors by choosing a fixed Riemannian metric in $\so$
\begin{equation}\lae{3.4.1} 
\chi=(\chi_{ij}(x)),
\end{equation}
and, for a given metric $g=(g_{ij}(t,x))$, we define
\begin{equation}
\f=\f(x,g_{ij})=\big(\frac{\det g_{ij}}{\det \chi_{ij}}\big)^\frac12
\end{equation}
such that the \ind{Einstein-Hilbert functional} $J$ in \fre{3.3.17} can be written in the form
\begin{equation}\lae{1.3.40}
J=\int_a^b\int_\Om\{\frac14G^{ij,kl}\dot g_{ij}\dot g_{kl}w^{-2}+(R-2\Lam)\}w\f\sqrt \chi.
\end{equation}
The Hamilton density $\mc H$ is then replaced by the function
\begin{equation}\lae{1.4.4}
H=\{\f^{-1}G_{ij,kl}\pi^{ij}\pi^{kl}-(R-2\Lam)\f\}w,
\end{equation}
where now
\begin{equation}
\pi^{ij}=-\f G^{ij,kl}h_{kl} 
\end{equation}
and
\begin{equation}\lae{1.4.6}
h_{ij}=-\f^{-1}G_{ij,kl}\pi^{kl}.
\end{equation}
The effective Hamiltonian is of course
\begin{equation}
w^{-1}H.
\end{equation}
Fortunately, we can, at least locally, assume
\begin{equation}
w=1
\end{equation}
by choosing an appropriate coordinate system: Let $(t_0,x_0)\in N$ be an arbitrary point, then consider the Cauchy hypersurface
\begin{equation}
M(t_0)=\{t_0\}\times \so
\end{equation}
and look at a tubular neighbourhood of $M(t_0)$, i.e., we define new coordinates $(t,x^i)$, where $(x^i)$ are coordinate for $\so$ near $x_0$ and $t$ is the signed Lorentzian distance to $M(t_0)$ such that the points
\begin{equation}
(0,x^i)\in M(t_0).
\end{equation}
The Lorentzian metric of the ambient space then has the form
\begin{equation}
d\bar s^2=-dt^2+g_{ij} dx^idx^j.
\end{equation}

Secondly, we use the same model as in \cite[Section 3]{cg:qgravity}: The Riemannian metrics $g_{ij}(t,\cdot)$ are elements of the bundle $T^{0,2}(\so)$. Denote by \inds{$E$} the \ind{fiber bundle} with base $\so$ where the fibers consist of the Riemannian metrics $(g_{ij})$. We shall consider each fiber to be a Lorentzian manifold equipped with the DeWitt metric. Each fiber $F$ has dimension 
\begin{equation}
\dim F=\frac{n(n+1)}2\equiv m+1.
\end{equation}
Let $(\xi^a)$, $0\le a\le m$, be  coordinates for a local trivialization such that 
\begin{equation}
g_{ij}(x,\xi^a)
\end{equation}
is a local embedding. The \ind{DeWitt metric} is then expressed as
\begin{equation}
G_{ab}=G^{ij,kl}g_{ij,a}g_{kl,b},
\end{equation}
where a comma indicates partial differentiation.  In the new coordinate system the curves 
\begin{equation}
t\ra g_{ij}(t,x)
\end{equation}
can be written in the form
\begin{equation}
t\ra \xi^a(t,x)
\end{equation}
and we infer
\begin{equation}
G^{ij,kl}\dot g_{ij}\dot g_{kl}=G_{ab}\dot\xi^a\dot\xi^b. 
\end{equation}
Hence, we can express \re{1.3.37} as
\begin{equation}\lae{1.3.49}
J=\int_a^b\int_\Om  \{\tfrac14 G_{ab}\dot\xi^a\dot\xi^b\f+(R-2\Lam)\f\},
\end{equation}
where we now refrain from writing down the density $\sqrt\chi$ explicitly, since it does not depend on $(g_{ij})$ and therefore should not be part of the Legendre transformation.  We also emphasize that we are now working in the gauge $w=1$.
Denoting the Lagrangian \tit{function} in \re{1.3.49} by $L$, we define 
\begin{equation}
\pi_a= \pde L{\dot\xi^a}=\f G_{ab}\frac1{2 }\dot\xi^b
\end{equation}
and we obtain for the Hamiltonian function $H$ 
\begin{equation}
\begin{aligned}
 H&=\dot\xi^a\pde L{\dot\xi^a}-L\\
&=\f G_{ab}\big(\frac1{2 }\dot\xi^a\big)\big(\frac1{2 }\dot\xi^b\big)  - (R-2\Lam)\f \\
&=\f^{-1}G^{ab}\pi_a\pi_b  - (R-2\Lam)\f ,
\end{aligned}
\end{equation}
where $G^{ab}$ is the inverse metric.

The fibers equipped with the metric
\begin{equation}\lae{3.4.16}
(\f G_{ab})
\end{equation}
are then globally hyperbolic Lorentzian manifolds as we proved in \cite[Theorem 1.4.2]{cg:qgravity-book}. In the fibers we can introduce new coordinates $(\xi^a)=(\xi^0,\xi^A)$, $0\le a\le m$, and $1\le A\le m$, such that
\begin{equation}
\tau\equiv \xi^0=\log\f
\end{equation}
and $(\xi^A)$ are coordinates for the hypersurface
\begin{equation}
M=\{\f=1\}=\{\tau=0\}.
\end{equation}
The Lorentzian metric in the fibers can then be expressed in the form 
\begin{equation}\lae{1.4.7}
ds^2=-\frac{4(n-1)}n\f d\tau^2+\f G_{AB}d\xi^Ad\xi^B,
\end{equation}
\cf \cite[equ. (1.4.28]{cg:qgravity-book}, where we note that in that reference is a misprint, namely, the spatial part of the metric has an additional factor $\frac{4(n-1)}n$ which should be omitted. Defining a new time variable $\xi^0=t$ by setting 
\begin{equation}
\f=t^2,
\end{equation}
we infer
\begin{equation}\lae{3.26}
ds^2=-\frac{16(n-1)}n dt^2+\f G_{AB}d\xi^Ad\xi^B.
\end{equation}
The new metric $G_{AB}$ is independent of $t$. When we work in a local trivialization of the bundle $E$ the coordinates $\xi^A$ are independent of $x$ as well as the time coordinate $t$, \cf \cite[Lemma 1.4.4]{cg:qgravity-book}.

We can now quantize the Hamiltonian setting using the original variables $g_{ij}$ and $\pi^{ij}$. We consider the bundle $E$ equipped with the metric \re{1.4.7}, or equivalently,
\begin{equation}\lae{3.4.28}
(\f G^{ij,kl}),
\end{equation}
which is the \tit{covariant} form, in the fibers and with the Riemannian metric $\chi$ in $\so$. Furthermore, let\indexs{$C^\un_c(E)$}
\begin{equation}
C^\un_c(E)
\end{equation}
be the space of real valued smooth functions with compact support in $E$.

In the quantization process, where we choose $\hbar=1$, the variables $g_{ij}$ and $\pi^{ij}$ are then replaced by operators $\hat g_{ij}$ and $\hat\pi^{ij}$ acting in $C^\un_c(E)$ satisfying the \ind{commutation relations}
\begin{equation}
[\hat g_{ij},\hat\pi^{kl}]=i\de^{kl}_{ij},
\end{equation}
while all the other commutators vanish. These operators are realized by defining $\hat g_{ij}$ to be the \ind{multiplication operator}\indss{$\hat g_{ij}$}
\begin{equation}
\hat g_{ij} u=g_{ij}u
\end{equation}
and \inds{$\hat \pi^{ij}$} to be the \tit{functional} differentiation\index{functional differentiation}\indss{$\dde{}{g_{ij}}$}
\begin{equation}
\hat\pi^{ij}=\frac1i \dde{}{g_{ij}},
\end{equation}
i.e., if $u\in C^\un_c(E)$, then
\begin{equation}
\dde{u}{g_{ij}}
\end{equation}
is the \ind{Euler-Lagrange operator} of the functional 
\begin{equation}
\int_{\so}u\sqrt\chi\equiv\int_\so u.
\end{equation}
Hence, if $u$ only depends on $(x,g_{ij})$ and not on derivatives of the metric, then
\begin{equation}
\dde{u}{g_{ij}}=\pde u{g_{ij}}.
\end{equation}
Therefore, the transformed Hamiltonian \inds{$\hat H$} can be looked at as the hyperbolic differential operator
\begin{equation}\lae{3.4.36}
\hat H=-\D-(R-2\Lam)\f,
\end{equation}
where $\D$ is the Laplacian of the metric in \re{3.4.28} acting on functions  
\begin{equation}
u=u(x,g_{ij}).
\end{equation}
We used this approach in \cite{cg:qgravity} to transform the Hamilton constraint to the \ind{Wheeler-DeWitt equation} 
\begin{equation}\lae{3.37}
\hat Hu=0\qq \text{in } E
\end{equation}
which can be solved with suitable Cauchy conditions. However, the above hyperbolic equation can only be solved abstractly because of the scalar curvature term $R$, which makes any attempt to apply separation of variables techniques impossible. Therefore, we discard the Wheeler-DeWitt equation by ignoring the Hamilton constraint and quantize the Hamilton equations instead. This approach is certainly as justified as quantizing the Hamilton constraint, which takes only the normal Einstein equations into account, whereas the Hamilton equations  are equivalent to the tangential Einstein equations. Furthermore, the resulting hyperbolic equation will be independent of $R$ and we can apply separation of variables.

Following Dirac the Poisson brackets in \fre{3.3.32} are replaced by $\frac1i$ times the commutators in the quantization process since $\hbar=1$, i.e., we obtain
\begin{equation}
\{\pi^{ij},H\}\ra i[\hat H, \hat\pi^{ij}].
\end{equation}
Dropping the hats in the following to improve the readability equation \re{2.34} is then transformed to
\begin{equation}\lae{3.39}
\begin{aligned}
i g_{ij}[H,\pi^{ij}]=(\frac n2-2)(\abs A^2-H^2)\f+\frac n2(R-2\Lam)\f -R\f,
\end{aligned}
\end{equation}
where we note that now $w=1$. We have
\begin{equation}\lae{3.4.41}
\begin{aligned}
i[H,\pi^{ij}]&=[H,\dde{}{g_{ij}}]\\
&= [-\D,\dde{}{g_{ij}}]-[(R-2\Lam)\f,\dde{}{g_{ij}}],
\end{aligned}
\end{equation}
\cf \re{3.4.36}. We apply both sides to functions $u\in C^\un_c(E)$, where we additionally require
\begin{equation}
u=u(g_{ij}),
\end{equation}
i.e., $u$ does not explicitly depend on $x\in\so$. Hence, we deduce 
\begin{equation}\lae{3.4.42}
[-\D,\dde{}{g_{ij}}]u=[-\D,\pde{}{g_{ij}}]u=-R^{ij}_{\hp{ij},kl}u^{kl},
\end{equation}
because of the Ricci identities, where\indss{$R^{ij}_{\hp{ij},kl}$}
\begin{equation}
R^{ij}_{\hp{ij},kl}
\end{equation}
is the Ricci tensor of the fiber metric \re{3.4.28} and 
\begin{equation}
u^{kl}=\pde u{g_{kl}}
\end{equation}
is the gradient of $u$.

For the second commutator on the right-hand side of \re{3.4.41} we obtain
\begin{equation}\lae{3.4.45}
\begin{aligned}
-[(R-2\Lam)\f,\dde{}{g_{ij}}]u=-(R-2\Lam)\f\pde u{g_{ij}}+\dde{}{g_{ij}}\{(R-2\Lam)u\f\},
\end{aligned}
\end{equation}
where the last term is the Euler-Lagrange operator of the functional
\begin{equation}
\begin{aligned}
\int_\so (R-2\Lam)u\f&\equiv \int_\so(R-2\Lam)u\f\sqrt\chi\\
&=\int_\so(R-2\Lam)u\sqrt g
\end{aligned}
\end{equation}
with respect to the variable $g_{ij}$, since the scalar curvature $R$ depends on the derivatives of $g_{ij}$. In view of \cite[equ. (1.4.84)]{cg:qgravity-book}  we have
\begin{equation}\lae{3.4.47} 
\begin{aligned}
\dde{}{g_{ij}}\{(R-2\Lam)u\f\}&=\frac12 (R-2\Lam) g^{ij}u\f-R^{ij}u\f\\
&+\f\{u_{;}^{\hp;ij}-\tilde\D u g^{ij}\}+(R-2\Lam)\f\pde u{g_{ij}},
\end{aligned}
\end{equation}
where the semicolon indicates covariant differentiation in $\so$ with respect to the metric $g_{ij}$, $\tilde \D$ is the corresponding Laplacian. We also note that
\begin{equation}
\begin{aligned}
D_ku&= \pde u{x^k}+\pde u{g_{ij}}\pde{g_{ij}}{x^k}\\
&=\pde u{x^k}=0.
\end{aligned}
\end{equation}
in Riemannian normal coordinates. Hence, we conclude that the operator on the left hand-side of equation \re{3.39} applied to $u$ is equal to
\begin{equation}
\frac n2(R-2\Lam)\f u-R \f u
\end{equation}
in $E$, since
\begin{equation}
g_{ij} R^{ij}_{\hp{ij},kl}=0,
\end{equation}
\cf \cite[equ. (1.4.89)]{cg:qgravity-book}. On the other hand, applying the right-hand side of \re{3.39} to $u$ we obtain
\begin{equation}
-(\frac n2-2) \D u+\frac n2 (R-2\Lam)\f u -R\f u,
\end{equation}
where the Laplacian is the Laplacian in the fibers, since
\begin{equation}
(\abs A^2-H^2)\f =\f^{-1}G_{ij,kl}\pi^{ij}\pi^{kl}\q\to \q -\D.
\end{equation}
Thus, we conclude
\begin{equation}
-(\frac n2-2)\D u=0
\end{equation}
in $E$, and we have proved the following theorem:
\bt
The quantization of equation \fre{2.34} leads to the hyperbolic equation
\begin{equation}\lae{3.54}
-\D u=0
\end{equation}
in $E$ provided $n\not= 4$ and $u\in C^\un_c(E)$ only depends on the fiber elements $g_{ij}$.
\et
To solve the equation \re{3.54} we first choose the Gaussian  coordinate system $(\xi^a)=(t,\xi^A)$ such that the metric has form as in \re{3.26}. Then, the hyperbolic equation can be expressed as
\begin{equation}
\frac n{16(n-1)}t^{-m}\pde{}{t}(t^m \pde ut)-t^{-2}\bar\D u=0,
\end{equation}
where $\bar\D$ is the Laplacian of the hypersurface
\begin{equation}
M=\{t=1\}.
\end{equation}
We shall try to use separation of variables by considering solutions $u$ which are products
\begin{equation}\lae{3.57}
u(t,\xi^A)=w(t)  v(\xi^A),
\end{equation}
where $v$ is a spatial eigenfunction, or eigendistribution, of the Laplacian $\bar\D$
\begin{equation}
-\bar\D v=\lam v
\end{equation}
and $w$ a temporal eigenfunction satisfying the ODE  
\begin{equation}\lae{3.59}   
\frac n{16(n-1)}t^{-m}\pde{}{t}(t^m \pde wt)+\lam t^{-2} w=0
\end{equation}
which can be looked at as an implicit eigenvalue equation. The function $u$ in \re{3.57} will then be a solution of \re{3.54}.

In the next sections we shall determine spatial and temporal eigendistributions by assuming 
\begin{equation}
\so=\R[n]
\end{equation}
equipped with the Euclidean metric. The dimension $n$ is then merely supposed to satisfy $n\ge 3$, though, of course, the equation \re{3.54} additionally requires $n\not= 4$.

\section{Spatial eigenfunctions in $M$} \las{4}

The hypersurface
\begin{equation}
M=\{\f=1\}
\end{equation}
can be considered to be a sub bundle of $E$, where each fiber $M(x)$ is a hypersurface in the fiber $F(x)$ of $E$. We shall use the same notation $M$ for the  sub bundle as well as for the hypersurface and in general we shall omit the reference to the base point $x\in \so$. Furthermore, we specify the metric $\chi_{ij}\in T^{0,2}(\so)$, which we used to define $\f$, to be equal to the Euclidean metric such that in Euclidean coordinates
\begin{equation}
\f^2=\frac{\det g_{ij}}{\det \de_{ij}}=\det g_{ij}.
\end{equation}
Then, it is well-known that each $M(x)$ with the induced metric $(G_{AB})$ is a symmetric space, namely, it is isometric to the coset space
\begin{equation}\lae{4.3}
G/K=SL(n,\R[])/SO(n),
\end{equation}
\cf \cite[equ.(5.17), p. 1123]{dewitt:gravity} and \cite[p. 3]{jorgenson:book}. The eigenfunctions in symmetric spaces, and especially of the coset space in \re{4.3}, are well-known, they are the so-called \tit{spherical functions}. One can also define a Fourier transformation for functions in $L^2(G/K)$ and prove a Plancherel formula, similar to the Euclidean case, \cf \cite[Chapter III]{helgason:ga}. Also similar to the Euclidean case we shall use the Fourier kernel to define the eigenfunctions, or eigendistributions, since the spherical functions, because of their symmetry properties, are not specific enough to represent the elementary gravitons corresponding to the diagonal metric variables $g_{ii}$, $1\le i\le n-1$. Recall that from the $n$ diagonal coefficients of a metric only $n-1$ are independent because of the assumption
\begin{equation}
\det g_{ij}=1
\end{equation}
which has to be satisfied by the elements of $M$.

But before we can define the eigenfunctions and analyze their properties, we have to recall some basic definitions and results of the theory of symmetric spaces. We shall mainly consider the coset space in \re{4.3} which will be the relevant space for our purpose. Its so-called \tit{quadratic model}, the naming of which will be obvious in the following, is the space of symmetric positive definite matrices in $\R[n]$ with determinant equal to $1$, i.e., the quadratic model of $G/K$ is identical to an arbitrary fiber $M(x)$ of the sub bundle  $M$ of $E$. Since the symmetric space $G/K$, as a Riemannian space, is isometric to its quadratic model, the eigenfunctions of the Laplacian in the respective spaces can be identified via the isometry.

Unless otherwise noted the symbol $X$ should denote the coset space $G/K$, where $G$ is the Lie group $SL(n,\R[])$ and $K=SO(n)$. The elements of $G$ will be referred to by $g,h,\ldots$, we shall also express the elements in $X$ by $x,y,\ldots$, and by a slight abuse of notation the elements of $M$ will also occasionally  be referred to by the symbol $g$, but always in the form $g_{ij}$.

The canonical isometry between the quadratic model $M$ and $X$ is given by the map
\begin{equation}\lae{4.5}
\begin{aligned}
\pi:G/K&\ra M\\
x=gk\in gK&\ra \pi(x)=gk(gk)^*=gg^*,
\end{aligned}
\end{equation}
where the star denotes the transpose, hence, the name quadratic model. For fixed $(g_{ij})\in M$, the action
\begin{equation}
[g](g_{ij})=g(g_{ij}) g^*, \qq g\in G,
\end{equation}
is an isometry in $M$, where $M$ is equipped with the metric
\begin{equation}
\tilde G^{ij,kl}=\frac 12\{g^{ik}g^{jl}+g^{il}g^{jk}\},
\end{equation}
and where
\begin{equation}
(g^{ij})=(g_{ij})^{-1},
\end{equation}
\cf \cite[equ. (1.4.46), p. 22]{cg:qgravity-book}. 

Let
\begin{equation}\lae{4.9}
G=NAK
\end{equation}
be an Iwasawa decomposition of $G$, where $N$ is the subgroup of unit upper triangle matrices, $A$ the abelian subgroup of diagonal matrices with strictly positive diagonal components and $K=SO(n)$. The corresponding Lie algebras are denoted by
\begin{equation}
\mf g, \mf n,\mf a\tup{\, and\, }\mf k.
\end{equation}
Here,
\begin{equation}
\begin{aligned}
\mf g&= \tup{real matrices with zero trace}\\
\mf n&=\tup{subspace of strictly upper triangle matrices with zero diagonal}\\
\mf a&=\tup{subspace of diagonal matrices with zero trace}\\
\mf k&=\tup{subspace of skew-symmetric matrices}.
\end{aligned}
\end{equation}
The Iwasawa decomposition is unique. When
\begin{equation}
g=nak
\end{equation}
we define the maps $n, A, k$ by
\begin{equation}\lae{4.13}
g=n(g)A(g)k(g).
\end{equation}
We also use the expression $\log A(g)$, where $\log$ is the matrix logarithm. In case of diagonal matrices
\begin{equation}
a=\diag(a_1,\dots, a_n)
\end{equation}
with positive entries
\begin{equation}
\log a=\diag(\log a_i),
\end{equation}
hence
\begin{equation}
A(g)=e^{\log A(g)}.
\end{equation}
Helgason uses the symbol $A(g)$ if $G$ decomposed as in \re{4.9} but uses the symbol $H(g)$ if
\begin{equation}
G=KAN
\end{equation}
which can be obtained by applying the isomorphism
\begin{equation}
g\ra g^{-1}.
\end{equation}
Because of the uniqueness
\begin{equation}
H(g)=A(g)^{-1},
\end{equation}
hence
\begin{equation}
\log H(g)=-\log A(g),
\end{equation}
\cf \cite[equs. (2),(3), p 198]{helgason:ga}.

Note that the functions we define in $G$ should also be defined in $G/K$, i.e., we would want that
\begin{equation}
A(g)=A(gK),
\end{equation}
which is indeed the case. If we used the Iwasawa decomposition $G=KAN$, then
\begin{equation}
H(g)=H(Kg)
\end{equation}
would be valid which would be useful if we considered the right coset space $K\sminus G$.

\br\lar{4.1}
(i) The Lie algebra $\mf a$ is a (n-1)-dimensional real algebra, which, as a vector space, is equipped with a natural real, symmetric scalar product, namely, the trace form
\begin{equation}
\spd{H_1}{H_2}=\tr (H_1H_2),\qq H_i\in\mf a.
\end{equation}

\cvm
(ii) Let $\mf a^*$ be the dual space of $\mf a$. Its elements will be denoted by Greek symbols, some of which have a special meaning in the literature. The linear forms are also called \tit{additive characters}.

\cvm
(iii) Let $\lam\in \mf a^*$, then there exists a unique matrix $H_\lam\in\mf a$ such that
\begin{equation}
\lam(H)=\spd{H_\lam}H\qq\A\, H\in \mf a.
\end{equation}
This definition allows to define a dual trace form in $\mf a^*$ by setting for $\lam,\mu\in \mf a^*$
\begin{equation}\lae{4.25}
\spd \lam \mu=\spd{H_\lam}{H_\mu}.
\end{equation}

\cvm
(iv) The Lie algebra $\mf g$ is a direct sum
\begin{equation}
\mf g=\mf n+\mf a+\mf k.
\end{equation}
Let $E_{ij}$, $1\le i<j\le n$, be the matrices with component $1$ in the entry $(i,j)$ and other components zero, then these matrices form a basis of $\mf n$. For $H\in \mf a$, $H=\diag (x_i)$, the Lie bracket in $\mf g$, which is simply the commutator, applied to $H$ and $E_{ij}$ yields
\begin{equation}
[H,E_{ij}]=(x_i-x_j)E_{ij}\q\A\, H\in\mf a.
\end{equation}
Hence, the $E_{ij}$ are the eigenvectors for the characters $\al_{ij}\in \mf a^*$ defined by
\begin{equation}
\al_{ij}(H)=x_i-x_j.
\end{equation}
Here, $E_{ij}$ is said to be an eigenvector of $\al_{ij}$, if
\begin{equation}
[H,E_{ij}]=\al_{ij}(H) E_{ij}\qq\A\,H\in \mf a.
\end{equation}
The eigenspace of $\al_{ij}$ is one-dimensional. The characters $\al_{ij}$ are called the \tit{relevant} characters, or the $(\mf a,\mf n)$ characters. They are also called the positive restricted roots. The set of these characters will be denoted by $\Si^+$.
We define
\begin{equation}
\tau=\sum_{\al\in\Si^+}\al
\end{equation}
and 
\begin{equation}
\rho=\frac12\tau.
\end{equation}
\er
\bl
Let $H=\diag(x_i)\in\mf a$ and define
\begin{equation}
\lam_i(H)=\sum_{k=1}^ix_k,\q\tup{for}\q1\le i\le n-1,
\end{equation}
then
\begin{equation}\lae{4.33}    
\rho=\sum_{i=1}^{n-1}\lam_i. 
\end{equation}
Furthermore,
\begin{equation}\lae{4.34}
\spd\rho\rho=\frac1{12}(n-1)^2n.
\end{equation}
\el
\bp
\cq{\re{4.33}} Follows from the definition of $\rho$ and $\tau$. For details see \cite[p.~84]{jorgenson:book}.

\cq{\re{4.34}} From \re{4.25} we obtain
\begin{equation}
\spd\rho\rho=\spd{H_\rho}{H_\rho}
\end{equation}
and the definition of $\rho$ implies
\begin{equation}
H_\rho=\frac12 H_\tau.
\end{equation}
on the other hand,
\begin{equation}
H_\tau=\sum_{i=1}^{n-1}C_i,
\end{equation}
where $C_i\in\mf a$ has $1$ in the first $i$ entries of the diagonal, $-i$ in the $(i+1)$-th entry and zero in the other entries. Furthermore,
\begin{equation}
\spd{C_i}{C_j}=0,\qq i\not= j,
\end{equation}
and
\begin{equation}
\spd{C_i}{C_i}=i^2+i,
\end{equation}
\cf \cite[p.~266]{jorgenson:book}. Hence, we conclude
\begin{equation}\lae{4.40}  
\spd\rho\rho=\frac14\sum_{i=1}^{n-1}(i^2+i)=\frac1{12}(n-1)^2n.
\end{equation}
\ep
\br\lar{4.3}
The eigenfunctions of the Laplacian will depend on the additive characters. The above characters $\al_{ij}$, $1\le i<j\le n$, will represent the \tit{elementary gravitons} stemming from the degrees of freedom in choosing the coordinates
\begin{equation}
g_{ij},\qq 1\le i<j\le n,
\end{equation} 
of a metric tensor. The diagonal elements offer in general additional $n$ degrees of freedom, but in our case, where we  consider metrics satisfying
\begin{equation}
\det g_{ij}=1,
\end{equation}
only $(n-1)$  diagonal components can be freely chosen, and we shall choose the first $(n-1)$  entries, namely,
\begin{equation}
g_{ii},\qq1\le i\le n-1.
\end{equation}
The corresponding additive characters are named $\al_i, 1\le i\le n-1$, and are defined by
\begin{equation}
\al_i(H)=h_i,
\end{equation}
if
\begin{equation}
H=\diag(h_1,\ldots, h_n).
\end{equation}
The characters $\al_i$, $1\le i\le n-1$, and $\al_{ij}$ $1\le i<j\le n$, will represent the $\frac{(n+2)(n-1)}2$ \tit{elementary gravitons} at the character level. We shall normalize the characters by defining
\begin{equation}\lae{4.46}
\tilde\al_i=\norm{H_{\al_i}}^{-1}\al_i
\end{equation}
and
\begin{equation}\lae{4.47}
\tilde\al_{ij}=\norm{H_{\al_{ij}}}^{-1}\al_{ij}
\end{equation}
such that the normalized characters have unit norm, \cf \re{4.25}.
\er
\bd\lae{4.4}
Let $\lam\in\mf a^*$, then we define the \tit{spherical function}
\begin{equation}
\f_\lam (g)=\int_K e^{(i\lam+\rho)\log A(kg)}dk, \qq g\in G,
\end{equation}
where the Haar measure $dk$ is normalized such that $K$ has  measure $1$, and where $G=NAK$.
\ed
Observe, that 
\begin{equation}
\f_\lam(g)=\f_\lam(gK),
\end{equation}
i.e., $\f_\lam$ can be lifted to $X=G/K$.

The Weyl chambers are the connected components of the set
\begin{equation}
\mf  a \setminus \uuu_{1\le i<j\le n}\al_{ij}^{-1}(0).
\end{equation}
They consist of diagonal matrices having distinct eigenvalues. The Weyl chamber $\mf a_+$, defined by
\begin{equation}
\mf a_+=\set{H\in\mf a}{\al_{ij}(h)>0, \q 1\le i<j\le n},
\end{equation}
is called the positive Weyl chamber and the elements $H\in\mf a_+$, $H=\diag(h_i)$, satisfy
\begin{equation}
h_1>h_2>\cdots>h_n.
\end{equation}
Let $M$ \resp $M'$ be the centralizer \resp normalizer  of $\mf a$ in $K$, then
\begin{equation}
W=M'/M
\end{equation}
is the Weyl group which acts simply transitive on the Weyl chambers. The Weyl group can be identified with the group $S_n$ of permutations in our case, i.e., if $s\in W$ and $H=\diag(h_i)\in \mf a$, then
\begin{equation}
s\cdot H=\diag(h_{s(i)}).
\end{equation}
The subgroup $M$ consists of the diagonal matrices $\diag(\e_i)$ with $\abs{\e_i}=1$.

Let $B$ be the homogeneous space
\begin{equation}
B=K/M,
\end{equation}
then $B$ is a compact Riemannian space with a $K$-invariant Riemannian metric, \cf \cite[Theorem 3.5, p.~203]{nomizu:II}. $B$ is known as the Furstenberg boundary of $X$ and the map
\begin{equation}
\begin{aligned}
\f: B\times A&\ra X\\
(kM,a)&\ra kaK
\end{aligned}
\end{equation}
is a differentiable, surjective map, while the restriction of $\f$ to
\begin{equation}
B\times A^+,\q A^+=\exp \mf a_+,
\end{equation}
is a diffeomorphism with an open, dense image; also,
\begin{equation}\lae{4.58}
X=K\ol{A^+}eK
\end{equation}
 \cf \cite[Prop. 1.4, p. 62]{helgason:ga}. 
If $x=gK$, $b=kM$ and $G=NAK$ we define
\begin{equation}
A(x,b)=A(gK,kM)=A(k^{-1}g),
\end{equation}
\cf \re{4.13}.

We are now ready to describe the Fourier theory and Plancherel formula, due to Harish-Chandra for $K$-bi-invariant functions, \cf \cite{Harish_Chandra_1958,Harish_Chandra_1958b} and \cite[p. 48]{Harish_Chandra_1966}, and by Helgason for arbitrary functions in $L^1(X)$ and $L^2(X)$, \cf \cite{helgason:fundamental} and \cite[Theorem 2.6]{helgason:dual}. The extension of the Fourier transform to the Schwartz space $\msc S(X)$ and its inversion is due to Eguchi and Okamato \cite{eguchi:schwartz}, this paper is only an announcement without proofs; the proofs are given in \cite{eguchi:eisenstein}. We follow the presentation in Helgason's book \cite[Chapter III]{helgason:ga}.

To simplify the expressions in the coming formulas the measures are normalized such that the total measures of compact spaces are $1$ and the Lebesgue measure in Euclidean space is normalized such that the Fourier transform and its inverse can be expressed by the simple formulas
\begin{equation}
\hat f(\xi)=\int_{\R[n]}f(x)e^{-i\spd\xi x}dx
\end{equation}
and
\begin{equation}
 f(x)=\int_{\R[n]}\hat f(\xi)e^{i\spd\xi x}d\xi.
\end{equation}
The Fourier transform for functions $f\in C^\un_c(X,\Cc)$ is then defined by
\begin{equation}\lae{4.62}
\hat f(\lam,b)=\int_Xf(x)e^{(-i\lam +\rho)\log A(x,b)}dx
\end{equation}
for $\lam\in\mf a$ and $b\in B$, or, if we define
\begin{equation}
e_{\lam,b}(x)=e^{(i\lam+\rho)\log A(x,b)},
\end{equation}
by
\begin{equation}
\hat f(\lam,b)=\int_Xf(x)\ol e_{\lam,b}(x)dx.
\end{equation}
The functions $e_{\lam,b}$ are real analytic in $x$ and are eigenfunctions of the Laplacian, \cf \cite[Prop. 3.14, p. 99]{helgason:ga},
\begin{equation}
-\D e_{\lam,b}=(\abs \lam^2+\abs\rho^2)e_{\lam,b},
\end{equation}
where 
\begin{equation}
\abs \lam^2=\spd \lam\lam,
\end{equation}
 \cf \re{4.25}, and similarly for $\abs\rho^2$. We also denote the Fourier transform by $\mc F$ such that
 \begin{equation}
\mc F(f)=\hat f.
\end{equation}
Its inverse $\mc F^{-1}$ is defined in $R(\mc F)$ by
\begin{equation}
f(x)=\frac 1{\abs W}\int_B\int_{\mf a^*}\hat f(\lam,b)\abs{\mf c(\lam)}^{-2}d\lam db,
\end{equation}
where $\mf c(\lam)$ is Harish-Chandra's  $\mf c$-function and 
\begin{equation}
\abs W=\card W,
\end{equation}
the number of elements in $W$, in our case $\abs W=n!$.

As in the Euclidean case a Plancherel formula is valid, namely, citing from \cite[Theorem 1.5, p. 202]{helgason:ga}:
\bt
The Fourier transform $f(x)\ra \hat f(\lam,b)$, defined by \re{4.62}, extends to an isometry of $L^2(X)$ onto $L^2(\mf a_+^*\times B)$ (with measure $\abs{\mf c(\lam)}^{-2}d\lam db$ on $\mf a_+^*\times B$). Moreover
\begin{equation}
\int_Xf_1(x)\bar f_2(x)dx=\frac1{\abs W}\int_{\mf a^*\times B}\hat f_1(\lam,b)\ol{\hat f_2}(\lam,b) \abs{\mf c(\lam)}^{-2}d\lam db.
\end{equation} 
\et
We shall consider the eigenfunctions $e_{\lam,b}$ as tempered distributions of the Schwartz space $\msc S(X)$ and shall use their Fourier transforms
\begin{equation}
\hat e_{\lam,b}=\de_{(\lam,b)}=\de_\lam \otimes \de_b
\end{equation}
as the spatial eigenfunctions of
\begin{equation}
\mc F(-\D)=m(\mu)=(\abs \mu^2+\abs\rho^2),
\end{equation}
which is a multiplication operator, in the next section.

\section{Fourier quantization}\las{5} 
The Fourier theory in $X=G/K$ which we described at the end of the preceding section uses the eigenfunctions
\begin{equation}
e_{\lam,b}(x)=e^{(i\lam+\rho)\log A(x,b)},\qq(\lam,b)\in \mf a^*\times B,
\end{equation}
as the Fourier kernel. The Fourier quantization in Euclidean space uses the Fourier transform of the Hamilton operator, or only the spatial part of Hamilton operator, which in our case is
\begin{equation}
-\D=-\D_M=-\D_X,
\end{equation}
and the Fourier transforms of the corresponding physically relevant eigenfunctions. If the Hamilton operator is the Euclidean Laplacian in $\R[n]$, then the spatial eigenfunctions would be
\begin{equation}
e^{i\spd\xi x}.
\end{equation}
Therefore, we consider the eigenfunctions $e_{\lam,b}$ as a starting point. As in the Euclidean case the $e_{\lam,b}$ are tempered distributions. We first need to extend the Fourier theory to the corresponding Schwartz space $\msc S(X)$ and its dual space $\msc S'(X)$, the space of tempered distributions.

Let $D(G)$ be the algebra of left invariant differential operators in $G$ and $\bar D(G)$ be the algebra of right invariant differential operators. Furthermore, let
\begin{equation}
\f_0=\fv{\f_\lam}{\lam=0}
\end{equation}
be the spherical function with parameter $\lam=0$. Then, $\f_0$ satisfies the following estimates
\begin{equation}
0<\f_0(a)\le\f_0(e)=1\qq\A\, a\in A,
\end{equation}
and
\begin{equation}
\f_0(a)\le c (1+\abs a)^de^{-\rho\log a},\qq a\in A^+,
\end{equation}
where
\begin{equation}
d=\card \Si^+
\end{equation}
the cardinality of the set of positive restricted roots. Here, we used the following definitions, for $g=k_1ak_2$  ($a\in A, k_1,k_2\in K$), \cf \fre{4.58},
\begin{equation}
\abs g=\abs a=\abs{\log a},
\end{equation}
and $c$ is a positive constant. 

The Schwartz space $\msc S(X)$ is then defined by
\bd
The Schwartz space $\msc S(G)$ is defined as the subspace of $C^\un(G,\Cc)$ the topology of which is given by the semi-norms
\begin{equation}
p_{l,D,E}(f)=\sup_{g\in G}(1+\abs g)^l\f_0(g)^{-1}\abs{(DEf)(g)}<\un
\end{equation}
for arbitrary $l\in\N$, $D\in D(G)$ and $E\in \bar D(G)$. The Schwartz space $\msc S(X)$ consists of those functions in $\msc S(G)$ which are right invariant under $K$.
\ed
The Fourier transform for $f\in \msc S(X)$ is then well defined
\begin{equation}
\hat f(\lam,b)=\int_X f(x)\ol e_{\lam,b}(x) dx.
\end{equation}
Integrating over $B$ we obtain  
\begin{equation}
\begin{aligned}
F(\lam)&=\int_B \hat f(\lam,b)db\\
&=\int_Xf(x)\int_Be^{(-i\lam+\rho)\log A(x,b)}db dx\\
&=\int_Xf(x) \f_{-\lam}(x)dx,
\end{aligned}
\end{equation}
\cf \cite[equ. (1.8)]{schlichtkrull:fourier} for the last inequality. Hence, we deduce
\bl
$F(\lam)$ satisfies
\begin{equation}\lae{5.12} 
F(s\cdot \lam)=F(\lam)\q\A\,s\in W.
\end{equation}
\el
\bp
The spherical function $\f_\lam$ has this property, \cf \cite[Theorem 5.2, p.~100]{jorgenson:book}.
\ep
Next, we define the Schwartz space $\msc S(\mf a^*\times B)$. Note that $\mf a^*$ is a Euclidean space, in our case $\mf a^*=\R[n-1]$, and $B=K/M$ is a compact Riemannian space. Hence, we define  the Schwartz space $\msc S(\mf a^*\times B)$ as follows, \cf \cite[Def.~2, p.~240]{eguchi:schwartz}:
\bd
Let $\msc S(\mf a^*\times B)$ denote the set of all functions $F\in C^\un(\mf a^*\times B,\Cc)$ which satisfy the following condition: for any natural numbers $l,m,q$
\begin{equation}
p_{l,m,q}(F)=\sup_{(\lam,b)\in\mf a^*\times B}(1+\abs \lam^2)^l\sum_{\abs\al\le m}\abs{(-\D_B+1)^qD^\al F}<\un,
\end{equation}
where $\al=(\al_1,\ldots,\al_r)$, $r=\dim \mf a^*$, is a multi-index
\begin{equation}
D^\al F=D_1^{\al_1}\cdots D_r^{\al_r}F
\end{equation}
are partial derivatives with respect to $\lam\in \mf a^*$, and $\D_B$ is the Laplacian in $B$. 
\ed
The semi-norms $p_{l,m,q}$ define a topology on $\msc S(\mf a^*\times B)$ with respect to which it is a Fr\'echet space.
\bt
The Fourier transform $\mc F$
\begin{equation}
\mc F:\msc S(X)\ra \msc S(\mf a^*\times B)
\end{equation}
is continuous and if we define
\begin{equation}
\hat{\msc S}(\mf a^*\times B)=\{F\in \msc S(\mf a^*\times B):F(\lam)=\int_BF(\lam,b)db \tup{ satisfies \re{5.12}}\},
\end{equation}
then
\begin{equation}
\mc F:\msc S(X)\ra \hat{\msc S}(\mf a^*\times B)
\end{equation}
is a linear topological isomorphism.
\et
\bp
Confer \cite[Theorem 4]{eguchi:schwartz} and \cite[Lemma 4.8.2 \& Theorem 4.8.3, p.~212]{eguchi:eisenstein} 
\ep
\br
Note that the measure in $\cfs$ is defined by
\begin{equation}
d\mu(\lam,b)=\frac 1{\abs W}\abs{\mf c(\lam)}^{-2}d\lam db
\end{equation}
and that the function 
\begin{equation}
\lam\ra \abs{\mf c(\lam)}^{-1}
\end{equation}
has slow growth, \cf \cite[Lemma 3.5, p.~91]{helgason:ga}.
\er
We can now define the Fourier quantization. Let $\msc S'(X)$ \resp $\hat{\msc S'}(\mf a^*\times B)$ be the dual spaces of tempered distributions, then
\begin{equation}
\mc F^{-1}:\hat{\msc S}(\mf a^*\times B)\ra \msc S(X)
\end{equation}
is continuous. Let ${(\mc F^{-1})}^*$ be the dual operator
\begin{equation}
{(\mc F^{-1})}^*:\msc S'(X) \ra \hat{\msc S'}(\mf a^*\times B)
\end{equation}
defined by
\begin{equation}
\spd\om{\mc F^{-1}(F(\lam,b))}=\spd{({\mc F^{-1})}^*\om}{F(\lam,b)}
\end{equation}
for $\om \in \mc S'(X)$ and $F\in \cfs$. Let 
\begin{equation}
F(\lam,b)=\hat f(\lam,b),\qq f\in \mc S(X), 
\end{equation}
then
\begin{equation}\lae{5.22}
\spd\om f=\spd{({\mc F^{-1})}^*\om}{\hat f}.
\end{equation}
Now, choose $\om=e_{\lam,b}$, where  $(\lam,b)\in \mf a^*\times B$ is  arbitrary but fixed, then
\begin{equation}
\spd \om f=\int_Xf(x)\ol e_{\lam,b}(x)dx=\hat f(\lam,b).
\end{equation}
Hence, we deduce
\begin{equation}
{(\mc F^{-1})}^*\om=\de_{(\lam,b)}=\de_\lam\otimes \de_b.
\end{equation}
\bl
Let $\om\in \mc S'(X)$ then we may call ${(\mc F^{-1})}^*\om$ to be the Fourier transform of $\om$
\begin{equation}
{(\mc F^{-1})}^*\om=\hat\om.
\end{equation}
\el
\bp
$\mc S(X)$ can be embedded in $\mc S'(X)$ be defining for $\om\in \msc S(X)$
\begin{equation}
\spd\om f=\int_X f\bar\om dx, \qq\A\, f\in \msc S(X).
\end{equation}
$\om$ is obviously an element of $\mc S'(X)$ and the embedding is antilinear. On the other hand, in view of the Plancherel formula, we have
\begin{equation}
\int_X f\bar\om dx=\int_{\mf a^*\times B}\hat f(\lam,b)\ol{\hat\om}(\lam,b)d\mu(\lam, b)
\end{equation}
and thus, because of \re{5.22},
\begin{equation}
\spd{{(\mc F^{-1})}^*\om}{\hat f}=\int_{\mf a^*\times B}\hat f\ol{\hat\om}d\mu(\lam,b).
\end{equation}
\ep
Looking at the Fourier transformed eigenfunctions
\begin{equation}
\hat e _{\lam,b}=\de_\lam\otimes \de_b
\end{equation}
it is obvious that the dependence on $b$ has to be eliminated, since there is neither a physical nor a mathematical motivation to distinguish between $e_{\lam,b}$ and $e_{\lam,b'}$. The first ansatz would be to integrate over $B$, i.e., we would consider the Fourier transform of
\begin{equation}
\int_Be_{\lam,b}db =\f_\lam
\end{equation}
which is equal to the Fourier transform of the spherical function $\f_\lam$, i.e.,
\begin{equation}
\hat\f_\lam=\de_\lam
\end{equation}
and it would act on the functions
\begin{equation}
F(\mu)=\int_B\hat f(\mu,b)db, \qq f\in \msc S(X).
\end{equation}
These functions satisfy the relation \re{5.12} which in turn implies
\begin{equation}
s\cdot \de_{\lam}=\de_{s^{-1}\cdot \lam}=\de_\lam\qq\A\, s\in W
\end{equation}
if $\lam$ was allowed to range in all of $\mf a^*$. Hence, we would have to restrict $\lam$ to the positive Weyl chamber $\mf a_+^*$, but then, we would not be able to define the eigenfunctions corresponding to the elementary gravitons $g_{ii}$, $2\le i\le n-1$, since the corresponding $\lam$ belong to different Weyl chambers, \cf \frr{4.3}.

Therefore, we pick a distinguished $b\in B$, namely,
\begin{equation}\lae{6.63}
b_0=eM,\qq e=\id\in K,
\end{equation}
and only consider the eigenfunctions $e_{\lam,b_0}$ with corresponding Fourier transforms
\begin{equation}
\de_{\lam}\equiv \de_\lam\otimes \de_{b_0}=\hat e_{\lam,b_0}, \qq \lam\in \mf a^*.
\end{equation}
Then we can prove:
\bl
Let $\de_\lam$ be defined as above, then for any $s\in W$ satisfying $s\cdot \lam\not=\lam$, there exists $F\in\hat{\msc S}(\mf a^*\times B)$ such that
\begin{equation}\lae{5.38}
\spd{\de_\lam}{F}=F(\lam,b_0)\not= F(s\cdot \lam,b_0)=\spd {\de_{s\cdot\lam}}F.
\end{equation}
\el
\bp
Let $\psi\in C^\un_c(\mf a^*)$ be a function satisfying
\begin{equation}
\psi(\lam)\not=\psi(s\cdot\lam)
\end{equation}
and choose $\h\in C^\un(B)$ with the properties
\begin{equation}
\h(b_0)=1
\end{equation}
and
\begin{equation}
\int_B \h db=0,
\end{equation}
then
\begin{equation}
F=\psi\h\in \cfs
\end{equation}
and satisfies \re{5.38}.
\ep
The Fourier transform of the Laplacian is a multiplication operator similar to the Euclidean case.
\bl
(i) Let $f\in \msc S (X)$, then
\begin{equation}
\mc F(-\D f)=m(\lam)\hat f(\lam,b)\in \cfs,
\end{equation}
where
\begin{equation}
m(\lam)=\abs \lam^2+\abs\rho^2,\qq\lam\in \mf a^*.
\end{equation}

(ii) Let $\om\in \msc S'(X)$, then $-\D\om$ is defined as usual
\begin{equation}\lae{5.45}
\spd{-\D\om}f=\spd\om{-\D f}
\end{equation}
and
\begin{equation}
\mc F(-\D\om)=m(\lam)\hat\om\in \cfd,
\end{equation}
where
\begin{equation}
\spd{m(\lam)\hat\om}{F(\lam,b)}=\spd{\hat\om}{m(\lam)F(\lam,b)}\qq\A\,F\in\cfs.
\end{equation}
\el
\bp
\cq{(i)} The result follows immediately by partial integration.

\cq{(ii)} From \re{5.22} and \re{5.45} we deduce
\begin{equation}
\begin{aligned}
\spd{-\D\om}f&=\spd{\hat\om}{\mc F(-\D f)}\\
&= \spd{\hat\om}{m(\lam)\hat f(\lam,b)}\\
&=\spd{m(\lam)\hat\om}{\hat f(\lam,b)}\\
&=\spd{\mc F(-\D\om)}{\hat f(\lam,b)}.
\end{aligned}
\end{equation}
\ep
Now, choosing
\begin{equation}
\om=e_{\lam,b_0},
\end{equation}
where $\lam\in\mf a^*$ is fixed, then
\begin{equation}
\mc F(-\D\om)=m(\mu)\hat \om=m(\mu)\de_\lam=m(\lam)\de_\lam,
\end{equation}
since
\begin{equation}
\spd{m(\mu)\de_\lam}{F(\mu,b)}=\spd{\de_\lam}{m(\mu)F(\mu,b)}=m(\lam)F(\lam,b_0).
\end{equation}
In \frr{4.3} we already identified the additive characters corresponding to the elementary gravitons, namely,
the characters
\begin{equation}
\al_{ij}, \q 1\le i<j\le n
\end{equation}
and
\begin{equation}
\al_i,\q 1\le i\le n-1.
\end{equation}
We shall now define the corresponding forms in $\mf a^*$ with arbitrary energy levels:
\bd\lad{5.9}
Let $\lam\in \R[]_+$ be arbitrary. Then we consider the characters
\begin{equation}
\lam \tilde{\al}_i\q \wed \q \lam \tilde{\al}_{ij},
\end{equation}
where we recall that the terms embellished by a tilde refer to the corresponding unit vectors. Then the eigenfunctions representing the elementary gravitons are $e_{\lam\tilde \al_i,b_0}$ and $e_{\lam\tilde\al_{ij},b_0}$.
\ed
The corresponding eigenvalue with respect to $-\D$ is $\abs\lam^2+\abs\rho^2$, where by a slight abuse of notation $\abs \lam^2 = \lam^2$ and $\abs \rho^2=\spd\rho\rho$. Note that $\abs\rho^2$ is always strictly positive, indeed
\begin{equation}
\abs{\rho(n)}^2\ge \abs{\rho(3)}^2 =1,
\end{equation}
 if $X=SL(n,\R[])/SO(n)$ and $n\ge3$, \cf \fre{4.40}. 

\section{Temporal eigenfunctions}
The  temporal eigenfunctions $w=w(t)$ have to satisfy the ODE \fre{3.59}. or equivalently, 
\begin{equation}\lae{6.1}
\Ddot w+m t^{-1}\dot w+\mu_0 t^{-2}w=0,
\end{equation}
where $\mu_0$ should be equal to
\begin{equation}
\mu_0=\frac{16(n-1)}n(\abs \lam^2+\abs\rho^2),
\end{equation}
and where $(\abs\lam^2+\abs\rho^2)$ is the eigenvalue of a spatial eigenfunction. 

To solve $\re{6.1}$ we make the ansatz 
\begin{equation}\lae{6.3}
w(t)=t^{-\frac{(m-1)}2}e^{i\mu\log t},\qq \mu>0,
\end{equation}
to obtain
\begin{equation}
\Ddot w +m t^{-1}\dot w+\mu_0 t^{-2}w=\{-\frac {(m-1)^2}4+\mu_0-\mu^2\}w.
\end{equation}
In order to choose $\mu$ such that the term in the braces vanishes, we have to ensure that
\begin{equation}\lae{6.5}
\mu_0-\frac {(m-1)^2}4>0.
\end{equation}
Now, the estimate
\begin{equation}
\mu_0-\frac {(m-1)^2}4\ge \frac{16(n-1)}n\rho^2-\frac{(m-1)^2}4
\end{equation}
is valid, where
\begin{equation}
\rho^2=\frac{(n-1)^2n}{12}
\end{equation}
and
\begin{equation}
m=\frac{(n-1)(n+2)}2.
\end{equation}
One can easily check that
\begin{equation}
\frac{16(n-1)}n\rho^2-\frac{(m-1)^2}4=
\begin{cases}
>0,&3\le n\le 16,\\
<0,&17\le n.
\end{cases}
\end{equation}
In case $n\ge 17$ and
\begin{equation}\lae{6.10}
\mu_0-\frac {(m-1)^2}4<0,
\end{equation}
we obtain the solution
\begin{equation}
w=c_1 t^{-\frac{m-1}2+\sqrt{\frac{(m-1)^2}4-\mu_0}}+ c_2t^{-\frac{m-1}2-\sqrt{\frac{(m-1)^2}4-\mu_0}},
\end{equation}
while for 
\begin{equation}\lae{6.12} 
\mu_0-\frac {(m-1)^2}4=0, 
\end{equation}
we get
\begin{equation}
w=c_1 t^{-\frac{m-1}2}+c_2 t^{-\frac{m-1}2} \log t.
\end{equation}
\br
In all three cases \re{6.5}, \re{6.10} and \re{6.12} we obtain two real independent solutions, which become unbounded, if the big bang ($t=0$) is approached and vanish, if $t$ goes to infinity. The two real solutions contained in \re{6.3}, which generate all possible temporal eigenfunctions, if $3\le n\le 16$, seem to be the physically relevant solutions.  
\er

\section{Conclusions}
Quantizing the Hamilton equations instead of the Hamilton constraint we obtained the simple equation
\begin{equation}
-\D u=0
\end{equation}
in the fiber bundle $E$ provided $n\not=4$, where the Laplacian is the Laplacian of the Wheeler-DeWitt metric in the fibers and where $u$ is a smooth function which is only defined in the fibers of $E$
\begin{equation}
u=u(g_{ij}(x)),\qq x\in \so=\R[n].
\end{equation}
Expressing then the fiber metric as in \fre{3.26} we can use separation of variables and write the solutions $u$ as products
\begin{equation}
u=w(t) v(g_{ij}(x,\xi^A)),
\end{equation}
where $g_{ij}(x,\xi^A)$ is a local trivialization of the sub bundle $M$ the fibers of which consists of the metrics $g_{ij}$ with unit determinant, or more precisely,
\begin{equation}
\frac{\det g_{ij}(x)}{\det\de_{ij}(x)}=1,
\end{equation}
where $\de_{ij}$ is the Euclidean metric. Using Euclidean coordinates in $\so$ we can identify the fibers $M(x)$ with the symmetric space
\begin{equation}
G/K=SL(n,\R[n])/SO(n).
\end{equation}
The Riemannian metric in $G/K$ is identical to the induced fiber metric of $M(x)$ such that the spatial eigenfunctions of the corresponding (spatial) Laplacians can also be identified. Due to the well-known Fourier theory in $G/K$ we choose the Fourier kernel elements
\begin{equation}
e_{\lam,b_0}(y)=e^{(i\lam+\rho) \log A(y,b_0)},\q \lam\in \mf a^*,
\end{equation}
where we used the Iwasawa decomposition $G=NAK$ and where $b_0$ is the distinguished point specified in \fre{6.63}. These smooth functions are tempered distributions and are eigenfunctions of the Laplacian
\begin{equation}
-\D e_{\lam,b_0}=(\abs \lam ^2+\abs \rho^2)e_{\lam,b_0},
\end{equation}
their Fourier transforms are Dirac measures
\begin{equation}
\hat e_{\lam,b_0}=\de_\lam\otimes \de_{b_0}.
\end{equation}
In Fourier space the Laplacian is a multiplication operator
\begin{equation}
\mc F(-\D f)=(\abs\lam^2+\abs\rho^2)\hat f(\lam,b)\qq\A\, f\in \msc S(G/K),
\end{equation}
where $\lam$ ranges in $\mf a^*$ and $b$ in the Furstenberg boundary $B$.

Let
\begin{equation}
\pi: G/K\ra M
\end{equation}
be the canonical isometry defined in \fre{4.5}, then the eigenfunctions $f$ in $G/K$ can be transformed to be eigenfunctions in the fibers of the sub bundle $M$ by defining
\begin{equation}
v(g_{ij}(x,\xi^A))=f(\pi^{-1}(g_{ij}(x,\xi^A)),
\end{equation}
i.e., 
\begin{equation}
e_{\lam,b_0}\circ \pi^{-1}\circ g_{ij}(x,\xi^A),\qq\lam\in \mf a^*,
\end{equation}
are the spatial eigenfunctions with eigenvalues $(\abs \lam^2 +\abs \rho^2)$. The eigenfunctions corresponding to the elementary gravitons we defined in \frd{5.9}. They are characterized by special characters $\al_i$, $1\le i\le n-1$, for the diagonal gravitons and $\al_{ij}$, $1\le i<j\le n$, for the off-diagonal gravitons.

The temporal eigenfunctions $w(t)$, which we defined in the previous section, have the properties that they become unbounded if $t\ra 0$ and they vanish, together with all derivatives, if $t\ra \un$.

Furthermore, if we consider $t<0$, then the functions
\begin{equation}
\tilde w(t)=w(-t),\qq t<0,
\end{equation}
also satisfy the ODE \fre{6.1} for $t<0$, i.e., they are also temporal eigenfunctions  if the light cone in $E$ is flipped.

Thus, we conclude
\bt
The quantum model we derived for gravity can be described by products of spatial and temporal eigenfunctions of corresponding self-adjoint operators with a continuous spectrum. The spatial eigenfunctions can be expressed as Dirac measures in Fourier space and the spatial Laplacian as a multiplication operator. The spatial eigenvalues are strictly positive
\begin{equation}
\abs\lam^2+\abs\rho^2\ge \abs\rho^2\ge\abs{\rho(3)}^2=1.
\end{equation}
Choosing $\lam=0$ we have a common ground state with smallest eigenvalue $\abs\rho^2$ which could be considered to be the source of the dark energy.

Furthermore, we have a big bang singularity in $t=0$. Since the same quantum model is also valid by switching from $t>0$ to $t<0$, with appropriate changes to the temporal eigenfunctions, one could argue that at the big bang two universes with different time orientations could have been created such that, in view of the CPT theorem, one was filled with matter and the other with anti-matter.
\et

%\backmatter
%\includepdf[pages=-]{/Users/claus/Documents/Scanned-Documents/}
\bibliographystyle{hamsplain}
%\bibliography{mrabbrev,publications}
\providecommand{\bysame}{\leavevmode\hbox to3em{\hrulefill}\thinspace}
\providecommand{\href}[2]{#2}

%\listoffigures

%\cleardoublepage

%\thispagestyle{empty}
%\closegraphsfile
\end{document}